# On the consistency in the adiabatic theorem and quantum geometric phase

Hua‐Zhong Li[*]

Advanced Research Center, Zhongshan University, Guangzhou, China 510275

Abstract

The recent discovery of inconsistency (MS inconsistency) in the adiabatic approximation is discussed. In particular, the so-called, inconsistency in Berry phase is analyzed. On the contrary to some authors, we found that the MS inconsistency persisted and the so-called inconsistency in Berry phase does not exist.

It is well-known that the adiabatic approximation in quantum mechanics is a very useful method in applications of quantum mechanics in various fields. It is very surprising that after eighty years of the establishment of the theorem, the self-consistency of this method was questioned in recently years[1]. The usual formal result is shown to produce inconsistency. The standard condition for the approximation is shown to be insufficient and does not guarantee the validity of the result of application. The question initiated by [1] excited a lot of discussions [2-7]. The answers are divergent. In the current literatures we can summarize the following questions in controversy i.e.

(1) Does the inconsistency (MS) pointed out in [1] really exist?[1-7]
(2) Is the standard condition for quantum adiabatic approximation not a sufficient condition?[1-3]
(3) Does there exist the so called inconsistency in quantum geometric phase besides the MS inconsistency?[4,5]

In the present note, we would like to present an answer to the third question from a point of view different from the existing literatures. The quantum adiabatic theorem can be expressed as follows. In an adiabatic process if the initial state of the system is an eigenstate of the Hamiltonian H(t) at t=0

$$H(t=0)|n(t=0)\rangle = E_0|n(t=0)\rangle \qquad (1)$$

then at final time t=T, the finial state wave function will be

$$|\psi_\tau(T)\rangle = \exp\left[-i\int_0^T E\left(\frac{t}{\tau}\right)dt\right]|n(T)\rangle - O\left(\frac{1}{\tau}\right) \quad (2)$$

where $\tau = \frac{1}{\varepsilon}$, $\varepsilon$ is the rate of change of $H(t)$, $\varepsilon = \frac{\partial H}{\partial t}$ or $\frac{\partial \lambda}{\partial t}$ when $H(t)$ is parameterized in terms of $\lambda(t)$. In the adiabatic limit $\varepsilon \to 0$ or $\tau \to \infty$,

$$O\left(\frac{1}{\tau}\right) \to 0 \quad (3)$$

$$|\psi(T)\rangle_{\tau \to \infty} = \exp\left[-i\int_0^T E\left(\frac{t}{\tau}\right)dt\right]|n(T)\rangle \quad (4)$$

---

[*] E-mail address: puslhz@mail.sysu.edu.cn

$|n(t)\rangle$ for $t=0$, $t=T$ etc. are instantaneous eigenstate of $H(t)$ at each instant.

When the rate of adiabatic changing $\varepsilon$ is small but finite, not infinitesimal. $O\left(\frac{1}{\tau}\right) \neq 0$. If one neglects this term of $O\left(\frac{1}{\tau}\right)$, the result is the so-called adiabatic approximation. In general it has been many decades that the condition for validity of adiabatic approximation is taken as

$$\frac{\left|\langle n(t)|\frac{\partial H}{\partial t}|m(t)\rangle\right|}{|E_n - E_m|^2} \ll 1 \quad n \neq m \quad (5)$$

This condition has been known for long time and was thought to be sufficient. Now, the recent investigation [1] argued that (5) is not a sufficient condition. Even though (5) is satisfied, the adiabatic approximation may not be correct and it may lead to inconsistency [1-7]. In particular, [4] argued if one take into account the contribution of Berry's phase the inconsistency can be resolved. But in this way it leads to a new inconsistency that Berry's phase vanishes. We would like to point out that in [4], the Berry's phase was taken twice in their proof so led to the mistaken conclusion. We give an argument to show that when the Berry's phase contribution is properly treated the MS inconsistency still exists and there is no inconsistency in Berry's phase.

Let us start with the usual convention of the phase of the instantaneous eigenfunction of the time dependent Hamiltonian. The eigenfunction of eigenequation of an operator is determined only up to a phase factor. It is usually understood that the arbitrary phase factor in the eigenfunction of Hamiltonian is taken by convention. The phase convention was discussed in [8]. Here we present a brief summary. The instantaneous eigen- equation for time dependent Hamiltonian H(t)

$$H(t)u_n(x,t) = E_n(t)u_n(x,t) \qquad (6)$$

for particular instants $t$ treated as constant parameter. The phase for $u_n$ can be shown to be

$$\gamma(t) = i\int \langle n|\dot{n}\rangle dt \qquad (7)$$

where $|n\rangle$ stands for $u_n(x,t)$,

$$\langle n|\dot{n}\rangle = \int u_n^* \frac{\partial}{\partial t} u_n dx ,$$

Taking a phase transformation

$$u'_n(x,t) = e^{i\gamma_n(t)} u_n(x,t) \qquad (8)$$

$u'$ is also an eigenfunction of (6).

Then (7) makes

$$\gamma'_n(t) = i \int \langle n' | \dot{n}' \rangle dt = 0 \tag{9}$$

for all *n*. So the phase of eigenfunction can be set equal to zero, $\gamma' = 0$. This is the phase convention which has been understood for many decades. In the adiabatic approximation the phase can be written explicitly

$$|\psi(t)\rangle = \exp[-i \int E_n(t') dt'] e^{i\gamma_n(t)} |n(t)\rangle \tag{10}$$

For a non-cyclic Hamiltonian the phase $e^{i\gamma_n(t)}$ in (10) can be absorbed into the phase of $u_n(x,t)$ in (6) and determined by the above convention. But for a cyclic processes this can not be done since for the cyclic processes

$$\gamma_n(c) = i \oint \langle n | \dot{n} \rangle dt \tag{11}$$

is non-trivial which is the geometrical phase discovered by Berry. This phase is known to be geometrical in character which can be made explicit by the parallel displacement of the state vector $|\psi\rangle$ for a close circuit in parameter space when H(t) is parameterized. The vector $|\psi\rangle$ obtained an overall turning angle after completing the closed path which is expressed as $e^{i\gamma_n(T)}$, T is the time for a cycle along closed path C.

When you keep explicitly $e^{i\gamma(t)}$ in (10), it means that the contribution of geometric phase is already taken into account. While if in (10) one omits $e^{i\gamma_n(t)}$ and consider it to be absorbed in $u_n(x,t)$ then for cyclic processes, one has to consider the cyclic parallel transport of $|n(t)\rangle$, and has contribution from (12) of $\gamma_n(c)$.

$$\langle n(o)|n(T)\rangle = e^{i\gamma(c)} \tag{12}$$

Following [4], in the quantum adiabatic approximation

$$|\psi(t)\rangle \approx e^{-i\int E_n(t'')dt'} e^{i\gamma_n(t)}|n(t)\rangle \tag{13}$$

For a unitary related system, from [1] in the adiabatic approximation the wavefunction reads

$$|\overline{\psi}(t)\rangle \approx e^{i\int E_n(t'')dt'}|n(0)\rangle \tag{14}$$

MS. [1] proved that (13)(14) gave

$$\langle \overline{\psi}(t)|\overline{\psi}(t)\rangle e^{i\gamma_n(t)}\langle n(0)|n(t)\rangle \neq 1 \tag{15}$$

This violates unitarity. [4] argued that for cyclic processes, using (12), the MS inconsistency (15) can be resolved, unitarity is preserved.

As we have said above, when in (13), $e^{i\gamma_n(t)}$ is written explicitly then for cyclic cases, (12) is already taken into account and (12) cannot be used. So MS in-consistency

$$\langle \overline{\psi}(T)|\overline{\psi}(T)\rangle e^{i\gamma_n(t)}\langle n(0)|n(T)\rangle \neq 1 \tag{16}$$

persists, where T is the time period for one cycle.

$$|\psi(T)\rangle = e^{-i\int_0^T E_n(t')dt} e^{i\gamma_n(c)}|n(T)\rangle \tag{17}$$

The claim in [4] was that the MS inconsistency suggests

$$\langle \psi(0)|\psi(T)\rangle = e^{-i\int_0^T E_n(t')dt'} \exp[i\gamma_n(c)]\langle n(0)|n(T)\rangle = e^{-i\int_0^T E_n(t')dt'} \tag{18}$$

In [4], it is concluded that "This implies that there is no observable Berry phase"[4]. We now see the argument that leads this conclusion is incorrect since in (18) the contribution of Berry phase was taken twice. The point is that the $\gamma_n(t)$ in eq. (13) and in (2) cannot be identified with the $\gamma_n(t)$ in eq. (1). They are of different meaning. $\gamma_n(t)$ in eq. (13), it is a usual integrable phase should have been chosen according to convention, while the $\gamma_n(t)$ in eq. (1) is not, it is an non-integral phase. If one follows the argument in [4] for cyclic evolution, (2) gives

$$\langle n(o)|n(T)\rangle \approx e^{-i\gamma_n(T)} \qquad (3)$$

Then the Berry phase $\gamma_n(T)$ would appear twice: once in eq. (1) already and the other one in eq. (13). This is what I regard as the error of double counting of Berry's phase in [4]. If one correctly treats the contribution of Berry's phase, Then (18) becomes

$$\langle \psi(0)|\psi(T)\rangle = e^{-i\int_0^T E_n(t')dt'} \exp[i\gamma_n(c)]$$

So there is no inconsistency in Berry phases. The Berry phases do not vanish.

Now let us discuss in more detail how [4] double counts the contribution of $\gamma_n(c)$. In [4] eq. (1), the phase factor $e^{i\gamma_n(t)}$ is written explicitly. This means that the phase of the instantaneous eigenstate $|n(t)\rangle$ has been fixed according to the convention. When the system evolutes a cyclic motion it generates $\gamma(c)$ the geometric phase. From the point of view of parallel displacement, transporting $\psi(t)$ or $|n(t)\rangle$ is transporting the same vector. When one keeps the factor $e^{i\gamma(t)}$ explicitly in eq. (1) of [4], it implies to transport $\psi(t)$. While eq. (10) in [4] implies transporting $|n(t)\rangle$. They produces same geometric phase. You can do it either once only. Once you keep explicitly $e^{i\gamma(t)}$ in eq. (1) in [4], there is no additional phase appearing in $|n(t)\rangle$. So for eq. (10) in [4], the phase difference in $|0\rangle$ and $|t\rangle$ can not be set equal to the geometric phase after t=T. The argument in [4] after eq. (15) suggests

$$\langle \psi(0)|\psi(T)\rangle = \exp[i\delta_n(T)]\exp(i\gamma_n(c))\langle n(0)|n(T)\rangle \qquad (20)$$

the authors claimed that

$$\langle n(0)|n(T)\rangle = \exp(-i\gamma_n(c)) \qquad (21)$$

such that

$$\langle \psi(0)|\psi(T)\rangle = \exp[i\delta_n(T)] \qquad (22)$$

It is concluded in [4] that "the initial and final state differ only by a dynamical phase and there is no observable Berry phase. Hence, a contradiction." Here we see that the Berry's phase was taken into account twice in [4]. One is in eq. (1) in [4] and another one is in the argument after eq. (14) in [4] using (21). This leads to the cancellation of $\gamma_n(c)$. As we have analyzed the phase $\gamma_n(t)$ cannot be counted twice according to the fundamental definition of the geometric phase. So the conclusion in [4] expressed as (22) is not correct.